# Transconductance and Coulomb blockade properties of in-plane grown carbon nanotube field effect transistors

Nan Ai[1], Onejae Sul[2], Milan Begliarbekov[1], Qiang Song[1], Kitu Kumar[2], Daniel S. Choi[3], Eui-Hyeok Yang[2], Stefan Strauf[1]

[1]Department of Physics & Engineering Physics, Stevens Institute of Technology
Castle Point on the Hudson, Hoboken, NJ 07030, USA

[2]Department of Mechanical Engineering, Stevens Institute of Technology
Castle Point on the Hudson, Hoboken, NJ 07030, USA

[3]Department of Materials Science & Engineering, University of Idaho
875 Perimeter Dr., Moscow, ID 83844, USA

## ABSTRACT

Single electron transistors (SETs) made from single wall carbon nanotubes (SWCNTs) are promising for quantum electronic devices operating with ultra-low power consumption and allow fundamental studies of electron transport. We report on SETs made by registered in-plane growth utilizing tailored nanoscale catalyst patterns and chemical vapor deposition. Metallic SWCNTs have been removed by an electrical burn-in technique and the common gate hysteresis was removed using PMMA and baking, leading to field effect transistors with large on/off ratios up to $10^5$. Further segmentation into 200 nm short semiconducting SWCNT devices created quantum dots which display conductance oscillations in the Coulomb blockade regime. The demonstrated utilization of registered in-plane growth opens possibilities to create novel SET device geometries which are more complex, i.e. laterally ordered and scalable, as required for advanced quantum electronic devices.

Corresponding author's E-mail: strauf@stevens.edu

## 1. INTRODUCTION

Single electron transistors (SETs) are promising nanoelectronic devices operating with ultra-low power consumption and allow fundamental studies of electron transport phenomena in the quantum regime [1, 2]. Single walled carbon nanotubes (SWCNTs) are of particular interest to realize SET devices operating at elevated temperatures since they have extremely small diameters of about 1 nm and since they can be densely integrated. To create a SET from a quantum wire a pair of local tunnel barriers must be introduced into an individual SWCNT, such that in effect a quantum dot (QD) is formed. The first SWCNT-SET devices have been made by simply making source drains contacts of a few hundred nm separation, which provide the tunnel barriers [3]. Recently, several different techniques have been utilized to create SWCNT-SET devices operating at higher temperatures, such as nicking by an atomic force microscope tip [4, 5], buckling by placing SWCNTs over a local bottom gate [6], local chemical modification [7], by adding local metallic top gates between source and drain contacts [8, 9], or by aligning metallic CNTs acting as a top-gate over a semiconducting SWCNT [10, 11]. In all these approaches individual SWCNTs are typically dispersed from an aqueous solution, dry out at random locations over the underlying substrate, and source/drain electrodes are subsequently fabricated by locating the SWCNT. Better control over the nanotube location can be achieved using prepatterned electrodes and ac dielectrophoresis alignment [12, 13], but control down to the limit of individual SWCNTs is limited and the contact resistance is typically large. SWCNTs can also be directly grown in-plane between source and drain electrodes at precisely defined locations using chemical vapor deposition (CVD) [14, 15]. Applications of CVD-grown SWCNT devices include densely integrated field effect transistors (FETs) utilizing post-growth transfer of SWNTs onto flexible substrates leading to output currents up to 1 A [16], and nanoelectromechanical switches operating at high-frequencies [17]. However, only few studies report SWCNT-SET devices made by in-plane CVD growth [18, 19], which required either additional shadow evaporation of electrodes or a local focused ion beam (FIB) manipulation technique. Here we demonstrate that on-chip in-plane grown semiconducting SWCNT devices can be either operated as FETs with high on/off ratios approaching $10^5$ or as SETs displaying pronounced conductance oscillations, without the need for post-growth transfer, shadow evaporation of electrodes, or local FIB manipulation.

## 2. SAMPLE FABRICATION AND MEASUREMENT TECHNIQUES

The CNTs studied in this paper were grown by chemical vapor deposition at $850^{\circ}$ C using a $CH_4/Ar_2$ mixture with a 1:4 ratio. At these conditions CNTs grow in-plane starting from a nanoscale Ni catalyst tip and bridge ultimately the Au/Cr contact electrodes. Further details of the growth procedure are given in our previous publication [20]. Figure 1 (a) shows a scanning electron microscope (SEM) picture with an individual SWCNT bridging the 2 μm separated electrode pair. Electrical characterization was performed in two-terminal geometry using a sensitive current meter with a resolution of about 1 pA and two source meters to control the source-drain and back-gate voltages. Samples were held on the cold-finger of a Helium flow cryostat with base temperature of 4.7 K. To prevent electrostatic discharge, the 200 nm thick Au/Cr bond pads were wire bonded to a chip carrier using an electrically grounded wire bonder tip and by applying only mechanical force and heat (~100° C) instead of ultrasound.

## 3. RESULTS AND DISCUSSION

### 3.1 Transformation of CNT arrays into FET devices with high on-off ratios

In-plane grown arrays of SWCNTs display either metallic or semiconducting properties corresponding to their chirality, which cannot be perfectly controlled in the CVD growth process. Metallic CNTs give rise to a non-vanishing off-state current, which masks the intrinsic high on/off ratios of the semiconducting SWCNTs. Since the metallic CNTs cannot be fully avoided in the CVD growth, we have adapted a technique to controllably break metallic CNTs without affecting the semiconducting ones [21, 22]. Since the metallic tubes in mixed arrays are detrimental for the device performance, an improvement FET performance is expected.

Figure 2 (a) shows a plot of the source-drain current $I_{sd}$ versus source-drain voltage $V_{sd}$ sweeps. The back gate was held constant at $V_g$= +20 V to switch the p-type semiconducting CNTs into their off-state. This prevents current flow through semiconducting CNTs and protects them from burning or degrading in the electrical breakdown procedure. Several breakdown events are visible at higher voltages. In order to eliminate all the metallic CNTs, we increased $V_{sd}$ until the first metallic CNT broke. The whole breakdown procedure was monitored by plotting the real-time current value. Once the current starts to fall, we stopped increasing $V_{sd}$ to prevent further damage. The next breakdown attempt starts from zero bias voltage until another metallic CNT breaks. After each breakdown, a characteristic $I_{sd}$ versus $V_g$ sweep was performed to examine the on/off ratio improvement. Figure 2 (b) shows the FET transconductance characteristic after the first and last break-in event, while steps in between are not shown. The initial transconductance sweep in Figure 2 (b) top shows a rather small on/off ratio of about 2 under 100 mV source-drain bias. When all metallic CNTs were removed, the field effect behavior of the device is dramatically enhanced and FET on/off ratio improved 4 orders of magnitude to about to $10^4$. While it is possible to create devices with only one SWCNT between an electrode pair by further reducing the electrode size, we have initially investigated larger electrodes with typically 20-30 SWCNTs across the contact gap. Therefore, several semiconducting SWCNTs contribute after burn-in to the total current in the on-state which reaches 0.1 μA under 100 mV source-drain voltage. The on-state current varies linearly with the source-drain voltage and highest on-state currents of about 1 μA have been achieved at about 1V bias voltage (not shown), corresponding to on/off ratios approaching $10^5$. These values are among the highest reported for CNT FETs which are back gated on 300 nm $SiO_2$ [23]. We have therefore successfully transformed mixed CNT arrays into pure arrays of semiconducting SWCNTs with high on/off ratios. Further improvement can only be achieved by top gating with thin high-k dielectrics [24, 25].

## 3.2 Control of conductance hysteresis in SWCNT FETs

Despite these outstanding transport properties, the fabricated FET devices suffer from gate hysteresis effects [26]. The transconductance characteristics of the SWCNT-FETs are strongly affected by the sweep direction of the gate voltage, i.e. sweeping from negative to positive voltages or sweeping backwards. The hysteresis effects do not only depend on the $V_g$ sweep direction, but they also depend on the chosen sweep range and sweep time. The larger the sweep range the larger the hysteresis effect (data not shown here). As is well known, the hysteresis is caused by charge trapping within the local CNT environment and is absent on elevated CNTs bridging an air gap [27]. In particular, it is not only affected by moisture in ambient conditions but also by the $SiO_2$ surface-bound water attached to the silane groups proximal to the nanotubes [28]. Figure 3 (a) shows the hysteresis sweep of an uncapped device measured in ambient atmosphere recorded under a gate sweep range from -10 V to + 10 V. As can be seen, the FET on/off switching threshold voltage changes by more than 8V. This hysteresis makes it nearly impossible to operate the FET as a SET device in the Coulomb blockade regime.

In order to eliminate hysteresis, we developed a procedure following Kim *et al*. [28]. Samples have been capped by PMMA followed by baking on a hot plate at 150 °C for several days in ambient air to remove surface-bound water molecules. The PMMA furthermore effectively encapsulates the CNT from the environment making measurements reproducible. Figure 3 (b) shows a gate sweep of the same device capped by PMMA and baked out while held in air. The magnitude of hysteresis was reduced to about 4 V and the current in the on-state was identical. Figure 3 (c) demonstrates a further reduction of that gate hysteresis to about 2 V if devices were held in the cryostat under vacuum. We also observed that the sample is less affected by the sweep range. Even though the nanotubes were covered by PMMA, measurements show that the vacuum environment can further reduce the hysteresis effect, indicating that the PMMA layer acts as a permeable membrane. After cooling the device to 80 K we observed a near-zero hysteresis effect in the same device as shown in Figure 3 (d). Since residual mobile charges are frozen out at low temperature the local CNT environment is less affected by changing the applied gate voltage direction. In conclusion, the hysteresis was strongly reduced utilizing the PMMA coating and baking technique and by cooling the device down to liquid nitrogen temperatures.

### 3.3 Electrical characterization of SWCNT-FET devices operating as SETs

Single electron transistors represent a subset of field effect transistors in which the gate voltage $V_g$ is used to position individual energy levels between the Fermi levels of the source and drain contacts. Ideally, if the Fermi level is positioned in the energy gap, under the application of a small source-drain bias, no current flows through the device: a transport regime known as Coulomb blockade. In this regime, individual energy levels can be moved in and out of the transport window (as defined by the source-drain potential) giving rise to Coulomb oscillations, as shown in Figure 4d. In the Coulomb blockade regime, all states lying below the Fermi levels of both the source and the drain contacts are filled, giving rise to the single electron charging energy gap $U_0 = U_{N+1} - U_N = q^2/C_\Sigma$ (as shown in Figure 4a), where $C_\Sigma$ is the total capacitance of the island, given by, $C_\Sigma = C_{SOURCE} + C_{DRAIN} + C_{GATE}$ [1, 2, 30]. It should be further noted that a prerequisite for observing Coulomb oscillations is that the level spacing should be smaller than the thermal energy of the electron $\Delta E < k_B T$, consequently, cryogenic temperatures are frequently utilized to achieve the above condition.

In accordance with previous findings, SWCNTs being either metallic or semiconducting, or even multiwall CNTs can operate as an SET if the device temperature is low enough, i.e. lower than the quantized level spacing and the charging energy [1, 29, 30,]. In effect, the Schottky barriers at the contacts define the quantization along the wire direction. The shorter the length of the quantum wire defined by the electrode spacing the larger the quantized level spacing ΔE. In order to reduce the contact spacing by an order of magnitude from 2000 nm to 200 nm additional electrodes have been deposited over an individual in-plane grown SWCNT as shown in Figure 1 (b).

While this device still contains about 10 semiconducting SWCNTs across, the transport signatures are expected to be dominated by the SET behavior of the SWCNT shortened to 200 nm, since it has a much larger level spacing. As shown in Figure 4 (a), electron transport over the discrete QD states, which are spaced out by $\Delta E$, can be controlled by changing the transport window width ($E_{FS}$-$E_{FD}$), which is defined by the difference in Fermi levels between source and drain. The larger this energy difference, which scales linearly with $V_{SD}$, the larger the number of transport channels contributing to the observed current. The role of the gate voltage is to tune the QD density of states through the transport window defined by the source-drain voltage. As a result, conductance oscillations are observed in Figure 4 (b-d). With each additional electron located on the QD the electrostatic potential energy $U_0$ increases, which affects the single electron transport. Therefore, the conduction oscillations vanish with increasing source-drain voltage as shown in Figure 4 (b). This is in agreement with the expected closing of the Coulomb diamonds typically observed when plotting the conductance as a function of source-drain voltage and gate-voltage [1, 2]. For the larger transport window of 1.5 mV highlighted in Figure 4 (c), the discrete electron jumps appear more as a step function since the transport window is broad enough to allow several transport channels to contribute simultaneously. Thus, the current does not go down in between each step. At the smaller transport window highlighted in Figure 4 (d) the conductance oscillations appear more like discrete peaks, since the window is small enough that only individual energy levels contribute to the total current.

It is furthermore observed that the overall spacing of the Coulomb blockade peaks is not constant at fixed $V_{SD}$ but slightly anharmonic and that peaks appear from an underlying current background which does not fully vanish to zero at all back gate voltages. This effect can have several reasons: First, the particular device contains other 2 micron long semiconducting SWCNTs which can contribute to the total current. Second, there is still a residual hysteresis since this device was only annealed for 12 hrs. Thus the frozen charges can affect the harmonicity of the energy level spacing. And third, the 200 nm long segment is still long enough to allow formation of several QDs along the segment during the carrier freeze out which have a nonlinear interplay and level spacing. These problems can be overcome in principle by fabricating contact pairs with a reduced number of Ni catalyst tips such that statistically only one SWCNT remains after burn-in per electrode pair and by further reducing the contact spacing itself.

## 4. SUMMARY

We have successfully fabricated in-plane grown arrays of purely semiconducting SWCNTs by utilizing tailored nanoscale catalyst patterns, an optimized low-pressure CVD growth process, and an electrical burn-in technique. The common gate hysteresis was removed using PMMA baking and FET devices were demonstrated with on/off ratios up to $10^5$, comparable to the highest values reported for back gated CNT FETs. For segmentation of bridged SWCNTs into QDs, we have characterized a 200 nm short SWCNT segment and observed conductance oscillations in the Coulomb blockade regime at 5K. As a result, we demonstrated SET operation of a preregistered and in-plane grown semiconducting SWCNT. In contrast to most previous work on CNT-SET devices, which requires either post-growth transfer of CNTs or local FIB manipulation techniques, the demonstrated utilization of registered in-plane growth opens possibilities to create laterally ordered and scalable SET device geometries, as required for advanced quantum electronic devices.

**ACKNOWLEDGMENTS**

We would like to acknowledge support by the Air Force Office for Scientific Research (Award No. FA9550-08-1-0134). In addition, the research has been carried out in part at the Center for Functional Nanomaterials, Brookhaven National Laboratory, which is supported by the U.S. Department of Energy, Office of Basic Energy Sciences, under Contract No. DE-AC02-98CH10886. M.B. acknowledges financial support by NSF GK-12 Grant No. DGE-0742462 .

# Figure Captions

**Figure 1.** (a) Scanning electron micrograph of an individual catalyst grown in-plane CNT crossing two Au/Cr electrodes. (b) Fabricated SET structure with a 200 nm SWCNT segment acting as a quantum dot. The added smaller electrodes effectively reduce the CNT segment length from the 2 μm as grown over the electrode gap down to 200 nm.

**Figure 2.** (a) Plot of source-drain current $I_{sd}$ versus source-drain voltage $V_{sd}$ for the whole breakdown procedure. (b) Characteristic of source-drain current $I_{sd}$ versus gate voltage $V_g$ before and after breakdown procedure. The on-state current was decreased due to the elimination of metallic channels, but the on/off ratio was improved up to $10^4$.

**Figure 3.** Characterization of gate hysteresis. (a) Isd versus Vg in forward and backward sweep direction on the uncapped device. Same sweep for a PMMA coated device held in air at RT (b), under vacuum at RT (c), and at liquid nitrogen temperatures (d). All data are recorded with 100 mV source-drain bias.

**Figure 4.** Characterization of SET performance. (a) Energy diagram for Coulomb blockade illustrating the transport window ($E_{FS}$-$E_{FD}$), the QD energy spacing $\Delta E$ and potential energy $U_0$ which increases for each additional loaded electron. (b) $I_{sd}$ versus $V_g$ sweeps showing Coulomb oscillations which vanish at higher $V_{sd}$ bias. (b). Panels (c) and (d) highlight two traces taken at $V_{sd}$ = 1.5 mV (c) and $V_{sd}$ = 0.5 mV (d). All measurements are carried out at 5.2K.

**Figure 1**

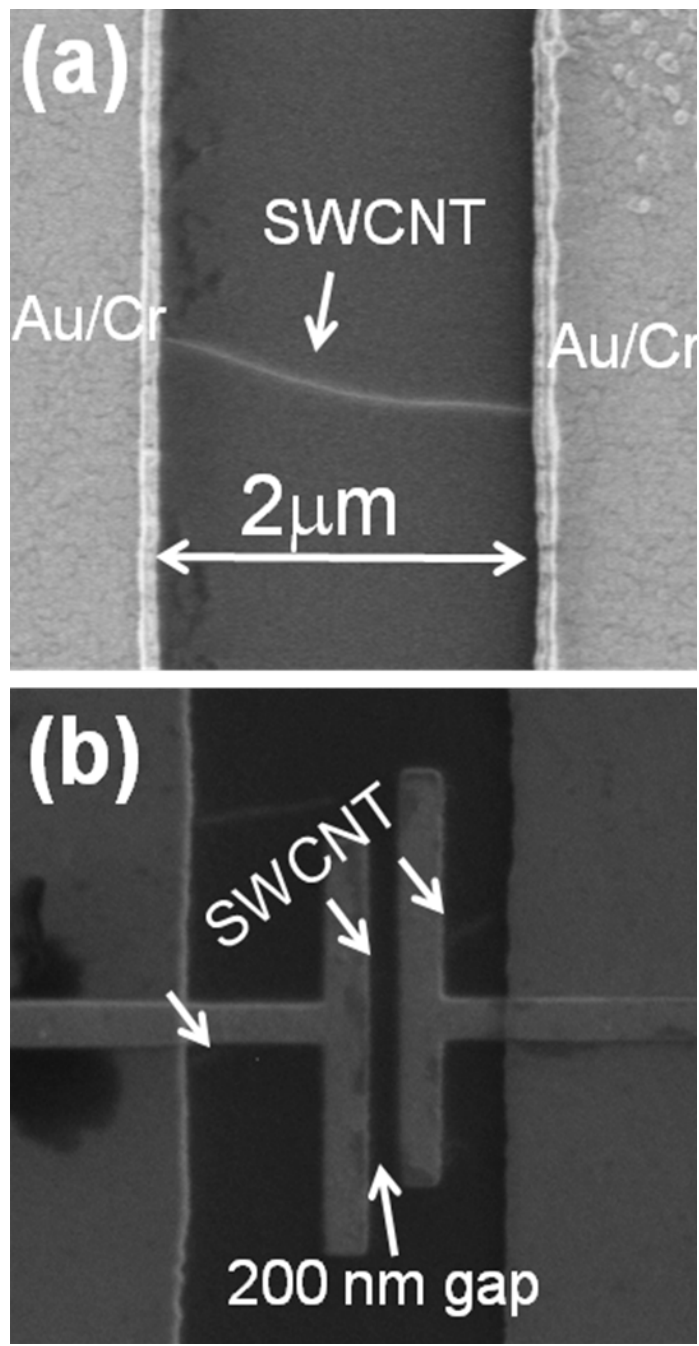

**Figure 2**

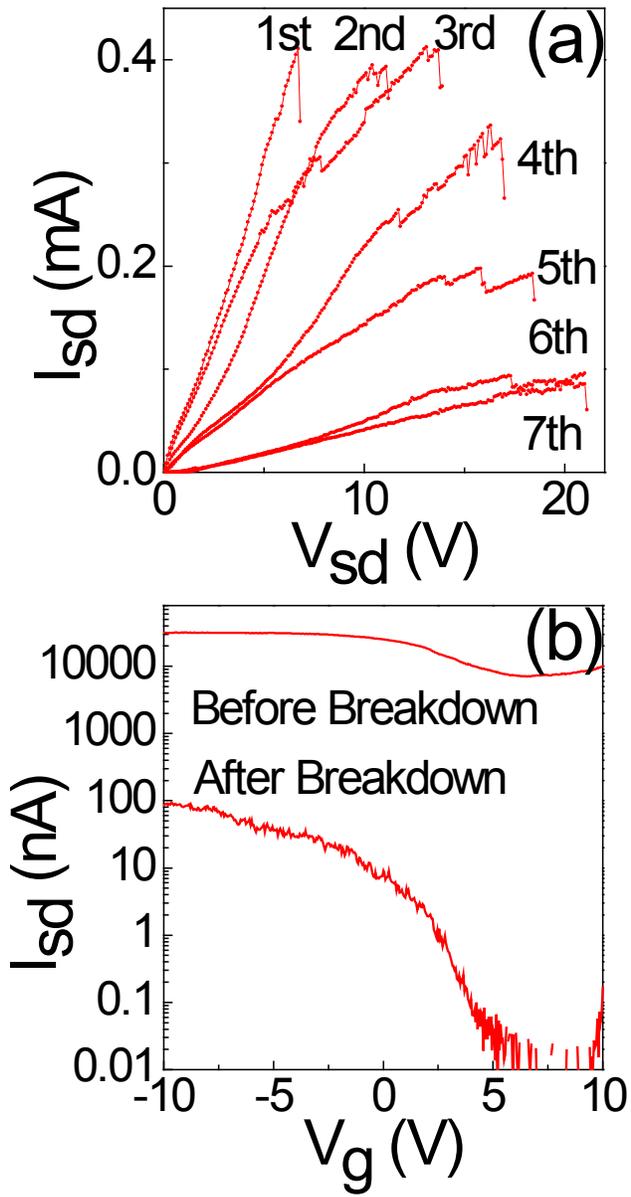

**Figure 3.**

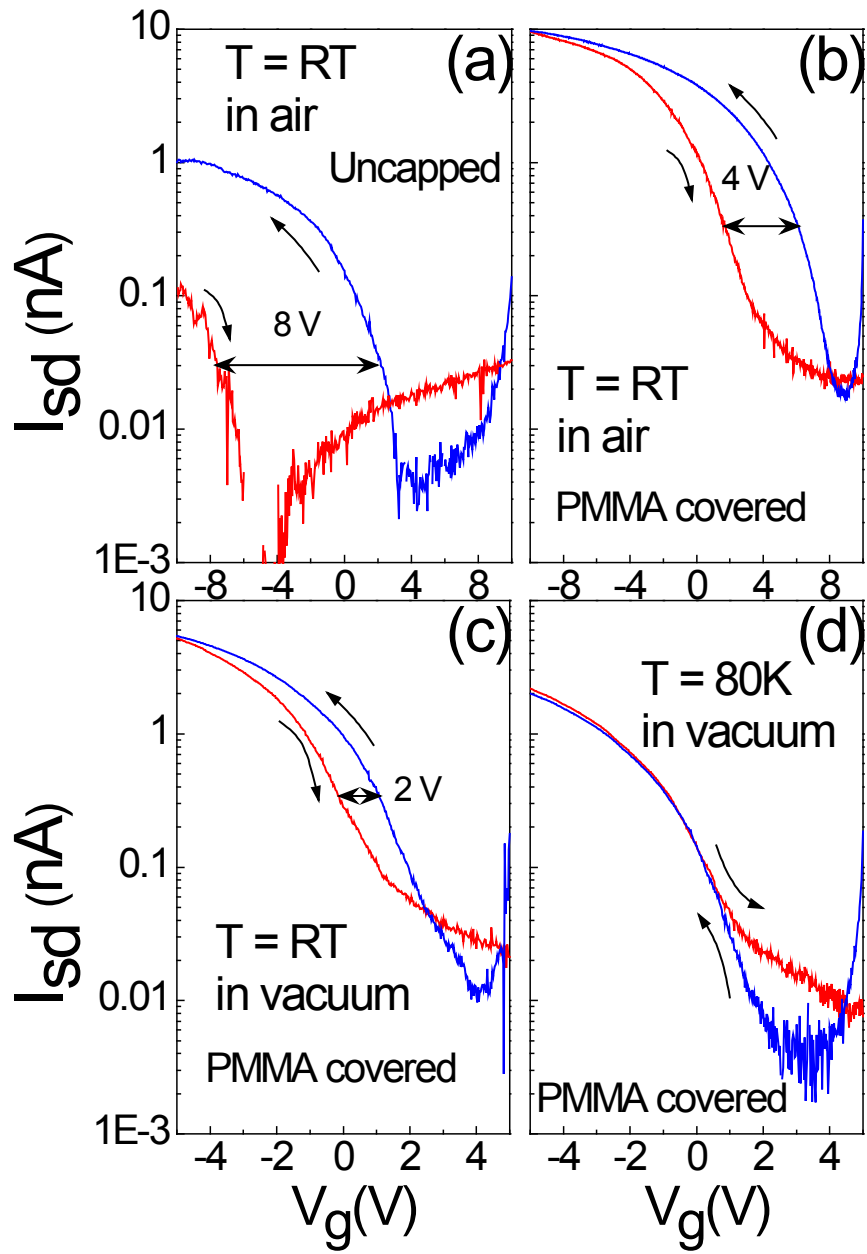

**Figure 4.**

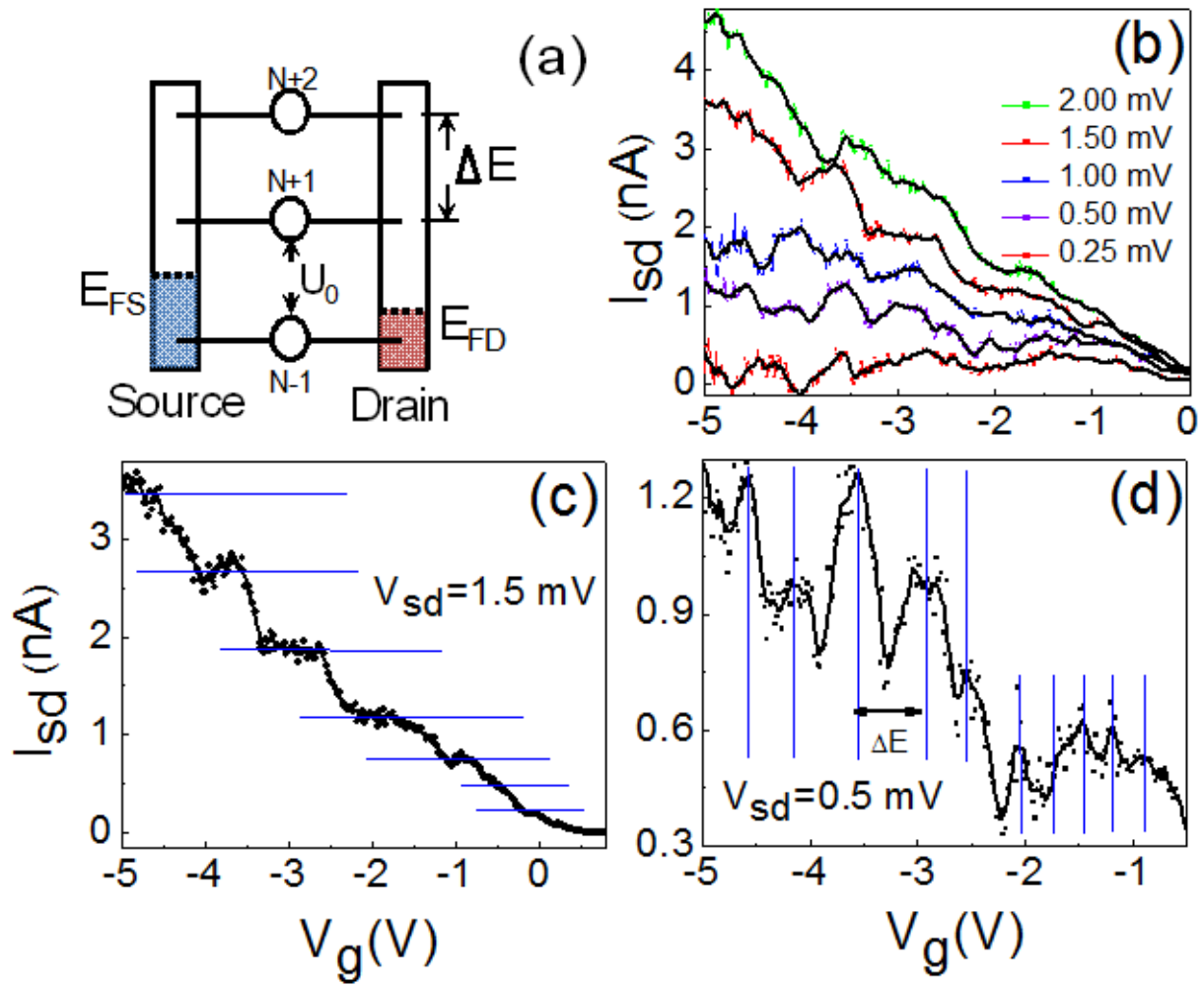